# Decentralized Gossip Mutual Learning (GML) for automatic head and neck tumor segmentation

Jingyun Chen*[a], Yading Yuan[a]

[a] Department of Radiation Oncology, Columbia University Irving Medical Center
622 W 168th St, New York, NY 10032, United States
*Email: jc6171@cumc.columbia.edu

## ABSTRACT

Federated learning (FL) has emerged as a promising strategy for collaboratively training complicated machine learning models from different medical centers without the need of data sharing. However, the traditional FL relies on a central server to orchestrate the global model training among clients. This makes it vulnerable to the failure of the model server. Meanwhile, the model trained based on the global data property may not yield the best performance on the local data of a particular site due to the variations of data characteristics among them. To address these limitations, we proposed Gossip Mutual Learning(GML), a decentralized collaborative learning framework that employs Gossip Protocol for direct peer-to-peer communication and encourages each site to optimize its local model by leveraging useful information from peers through mutual learning. On the task of tumor segmentation on PET/CT images using HECKTOR21 dataset with 223 cases from five clinical sites, we demonstrated GML could improve tumor segmentation performance in terms of Dice Similarity Coefficient (DSC) by 3.2%, 4.6% and 10.4% on site-specific testing cases as compared to three baseline methods: pooled training, FedAvg and individual training, respectively. We also showed GML has comparable generalization performance as pooled training and FedAvg when applying them on 78 cases from two out-of-sample sites where no case was used for model training. In our experimental setup, GML showcased a sixfold decrease in communication overhead compared to FedAvg, requiring only 16.67% of the total communication overhead.

## 1. INTRODUCTION

Federated learning (FL)[1] allows medical centers to collaborate on training deep-learning networks, such as tumor segmentation models, without sharing private data. However, traditional FL like FedAvg[1] relies on a central server to orchestrate the global model training among clients. This makes it vulnerable to the failure of the model server. Meanwhile, the model trained based on the global data property may not yield the best performance on the local data of a particular site due to the variations of data characteristics among them. This may discourage some medical centers from participating in FL. To address these limitations of FL, we proposed Gossip Mutual Learning (GML), a decentralized collaborative learning framework that employs Gossip Protocol for direct peer-to-peer communication and encourages each site to optimize its local model by leveraging useful information from peers through mutual learning.

## 2. MATERIALS AND METHODS

**Data**

This study employed 223 training cases in the 2021 HEad and neCK TumOR segmentation (HECKTOR21) challenge[2]. The tumor ground truth was annotated by multiple radiation oncologists, either directly on the CT of the PET/CT, or on a different CT registered to the PET/CT. These training cases were acquired across five different sites. In our work, three sites (IDs: CHMR, CHUM and CHUS) were used as participants for decentralized learning. For each of these three sites, 70% of cases were randomly selected for training, 10% were selected for validation, and 20% for local testing. The breakdown of case numbers is shown in Table 1. The remaining two sites (IDs: CHGJ and CHUP) were used entirely as out-of-sample testing cases.

**Segmentation model**

We used Scale Attention Network (SANet)[3,4] as the automatic tumor segmentation model in this work. SANet features a dynamic scale attention mechanism to incorporate low-lever details with high-level semantics from feature maps at different scales, thus, to better integrate information across different scales. It has demonstrated strong performance in different tasks such as head and neck tumor segmentation on PET/CT[5] and brain tumor segmentation on multi-parametric MRI images[6].

Table 1. Number of cases for the training sites

| Sites | Total cases | Training cases | Validation cases | Testing cases |
|---|---|---|---|---|
| CHMR | 18 | 11 | 3 | 4 |
| CHUM | 55 | 39 | 5 | 11 |
| CHUS | 72 | 51 | 7 | 14 |

**Gossip Mutual Learning**

Gossip Learning (GL)[7] has been proposed as a decentralized alternative to FL[8]. GL does not require a central server, as its model exchange is done in P2P style. However, conventional GL adopts weighted sum for model merging between two sites. This approach has limitations as the incoming model may not perform well on the receiver's data and may reduce the receiver's model performance through merging. To mitigate this problem, instead of directly merging models, we introduced a mutual learning[9] procedure to enable the information transfer by simultaneously training both the incoming and the local models on the local data for the task of tumor segmentation while aligning these two models.

Model alignment was achieved by introducing an additional function based on the Kullback-Leibler Divergence (KLD) between the predictions of these two models. As a tumor usually occupies a small region in the entire 3D images, we introduced a regional KLD (rKLD) loss to emphasize the agreement between these two models in the tumor region. For two models with predictions $P_1$ and $P_2$ correspondingly on voxels $x$, let $M$ represent the ground truth of segmentation, which is normally a manually identified tumor region, the rKLD loss is defined as:

$$rKLD(P_1, P_2|M) = \frac{\sum_{i,j,k}(P_1(x_{ijk}) \times \log\frac{P_1(x_{ijk})}{P_2(x_{ijk})} \times t_{ijk})}{\sum_{i,j,k} t_{ijk}} \quad (1)$$

where $\Sigma_{i,j,k}$ represents the sum over all voxel indices *(i, j, k)* for a 3D image, $t_{ijk} \in \{0,1\}$ is the true class of $x_{ijk}$, with $t_{ijk} = 1$ for tumor and $t_{ijk} = 0$ for background.

The general scheme of GML workflow is illustrated in Figure 1. Let the *i-th* participating site be associated with local dataset $D_i$, training sample size $\alpha_i$, and trainable model $W_i$. The major steps include:

1. Initialize all participating sites' local models with random weights, then train each model in parallel using its local dataset $D_i$ for a certain number of iterations, to allow "model warm-up".

2. Each participant maintains a random variable $S_i$ that follows a Bernoulli distribution with an activation probability $p_i$, i.e., $S_i \sim B(p_i)$. In this work, we set $p_i$ to be proportionate to $\alpha_i$. When $S_i = 1$ at the *t*-th iteration, the participant becomes a model sender, denoted as s. Another site is randomly selected from the rest participants, denoted as model receiver $r$

3. The sender will pause the local model training and send a connection request to the receiver. Once the request is accepted, $s$ will send its local model $W_s$ to $r$.

4. On the receiver, the local model $W_r$ is updated via regionalized mutual learning with $W_s$. The corresponding loss functions for $W_r$ and $W_s$ are set as:

$$L(W_r|W_s, M_r) = JD(P_r, M_r) + rKLD(P_r, P_s|M_r) \quad (2)$$

$$L(W_s|W_r, M_r) = JD(P_s, M_r) + rKLD(P_s, P_r|M_r) \quad (3)$$

where $P_s$ and $P_r$ are the predictions of models $W_s$ and $W_r$ on data $D_r$ correspondingly, $M_r$ is the ground truth of segmentation for $r$, $JD()$ is the Jaccard distance between prediction and ground truth[9], and $rKLD()$ is the rKLD loss defined in Equation (1).

5. If there are multiple incoming models, step 4 repeats until $Wr$ is updated sequentially with all the source models. Then the local model training resumes on $r$. For simplicity, in this work we send one incoming model at each GML iteration.

Since GML only updates the local model with incoming model(s), these local models will be different from each other at the end of GML training, yielding an individualized model in each participating site.

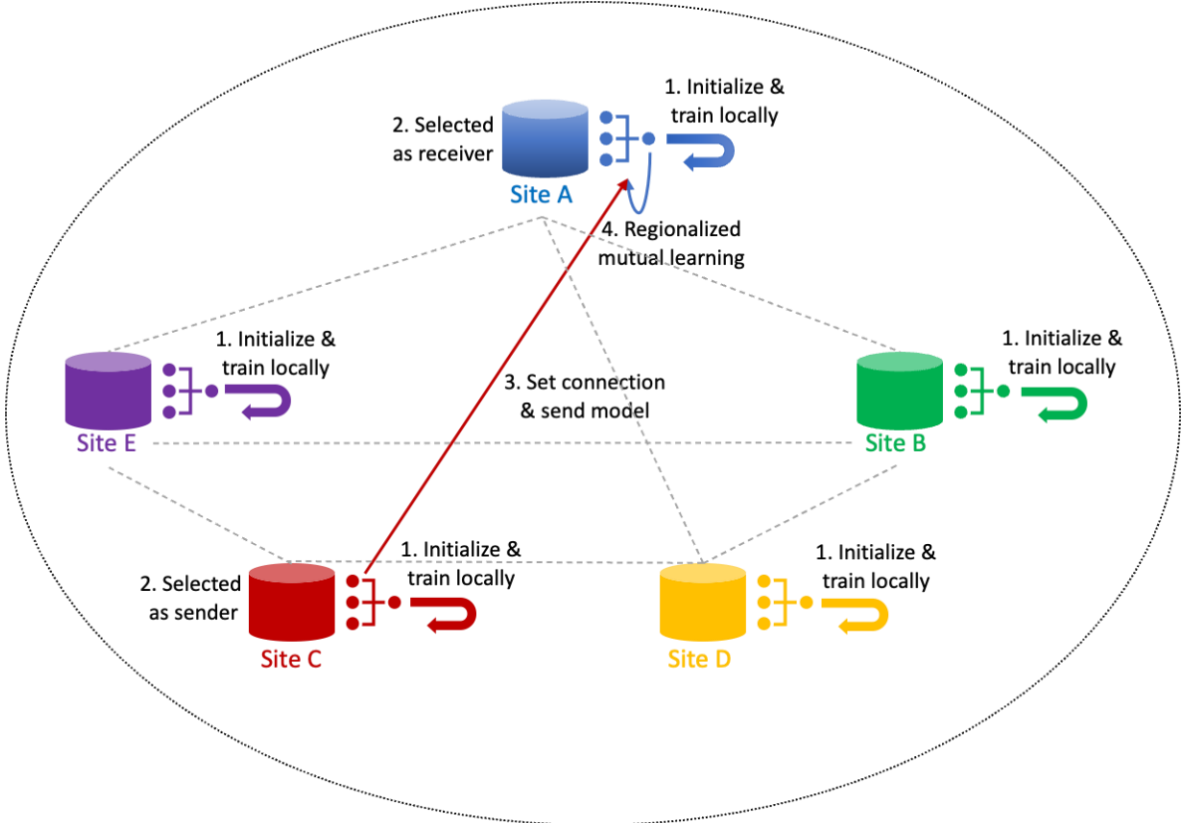

Fig 1. General scheme of GML. After initialization and warm-up, pairs of sender and receiver are randomly selected at each round. Then the sender transfers its model to the corresponding receiver. And the receiver performs regionalized mutual learning between incoming model and its local model.

**Baselines and evaluation**

We compare the proposed GML method against three baseline methods: 1) Pooled training: a global model is trained with all three sites' training data together. 2) Individual training: each site trains its local data individually without exchanging models. 3) FedAvg: federated learning among three sites with FedAvg aggregation[1]. We trained each method for 150 epochs. For GML and FedAvg, the first 50 epochs were used for warmup, followed by 100 epochs of normal training. To compare the performances of GML and baseline methods, we computed the mean Dice similarity coefficient (DSC) between model-predicted tumor segmentation and ground truth.

**Implementation**

Both the baseline and proposed methods were implemented with PyTorch, on Nvidia GTX 1080 TI GPUs with 11 GB memory. During FedAvg and GML, each site was assigned a separate GPU for training.

## 3. RESULTS

First, we compared GML and the bassline methods on **site-specific data** (Table 2). For Pooled training and FedAvg, the global models were tested. For GML and Individual training, the corresponding site-specific models were tested. The aggregated DSCs were computed by weighted average over three sites (CHMR, CHUM and CHUS), with weight equal to site's training case number. As shown in Table 2, GML has the highest aggregated DSCs, followed by Pooled training and FedAvg, with Individual training being lowest. GML also achieved consistently higher DSCs than Induvial training across all three sites, showing its potential to enhance local models on participating sites during decentralized learning. In comparison, FedAvg showed lower DSC than Individual training on CHUM, even though its aggregated DSC was higher.

Table 2. Mean DSCs on site-specific testing data. For Pooled and FedAvg, the global models were tested. For GML and Individual, the site-specific models were tested. The aggregated DSC is the weighted average of site-specific DSCs, with weights equal to site training case numbers.

| Methods | CHMR | CHUM | CHUS | Aggregated |
|---|---|---|---|---|
| Pooled | 0.7824 | 0.7009 | 0.7756 | 0.7482 |
| FedAvg | 0.7814 | 0.6677 | 0.7813 | 0.7382 |
| Individual | 0.5613 | 0.7120 | 0.7286 | 0.6992 |
| GML | 0.7279 | 0.7621 | 0.7920 | **0.7718** |

Second, we compared the mean DSCs of **site-specific models** by GML and Individual training, on both local testing and out-of-sample data (Table 3). The aggregated values are computed in the same way as Table 2. As shown in Table 3, GML consistently outperformed Individual training across all three sites, on both local testing and out-of-sample data. As expected, the improvement is also reflected in corresponding aggregated DSCs.

Table 3. Mean DSCs of site-specific models on local testing and out-of-sample data. The aggregated DSC is the weighted average of site-specific DSCs, with weights equal to site training case numbers.

| Data | Methods | CHMR | CHUM | CHUS | Aggregated |
|---|---|---|---|---|---|
| Local testing data | Individual | 0.5723 | 0.7433 | 0.7013 | 0.6994 |
| | GML | 0.6701 | 0.7313 | 0.7462 | **0.7301** |
| Out-of-sample data | Individual | 0.5501 | 0.7770 | 0.7108 | 0.7138 |
| | GML | 0.6077 | 0.7583 | 0.7459 | **0.7315** |

Finally, we compared the mean DSCs of **global models** by GML, FedAvg and Pooled training, on local testing and out-of-sample data. Since GML only maintains a local model on each participating site, we employed a bagging-type strategy to average the outputs of the site-specific models as the overall performance of GML when comparing it with the global models from the Pooled training and FedAvg. The results are shown in Table 4. On local testing data, GML showed higher DSC than both FedAvg and the Pooled training. On out-of-sample data, GML showed comparable performance as the Pooled training while outperforming FedAvg with only 1/6 communication overhead, because FedAvg requires aggregation and re-distribution of global model between the server and all the clients while GML only necessitates one communication between two clients at each round. These results demonstrate that the site-specific models trained with GML can be generalized to both in-sample and out-of-sample cases.

## 4. DISCUSSION

We presented a novel decentralized learning framework, namely GML, for automatic tumor segmentation on multi-modal medical images. Compared to the aggregation-based FL, the proposed GML has the following advantages: 1) GML is decentralized and more robust as it does not rely on a central server. 2) Rather than generating a global model with potentially suboptimal performance on local sites, GML aims to optimize local models on the participating sites. 3) GML's P2P communication is more efficient than aggregation-based FL (like FedAvg), as the later involves collecting from and returning the global mode to every site per learning round.

Table 4. Mean DSCs of global models on local testing and out-of-sample testing cases. For GML, the global model is generated through model ensemble of local models.

| Methods | Local testing data | Out-of-sample data |
|---|---|---|
| Pooled | 0.7482 | 0.7706 |
| FedAvg | 0.7382 | 0.7491 |
| GML | **0.7522** | **0.7596** |

## 5. CONCLUSIONS

GML is a robust and efficient framework for collaborative learning where data sharing is not required. By personalizing the local model, GML improves the performance of all participating sites, on both local and external data. With real clinical PET/CT from publicly available database, GML showed great potential in collaborative learning.

## ACKNOWLEDGEMENT

This work is supported by a research grant from Varian Medical Systems (Palo Alto, CA, USA), UL1TR001433 from the National Center for Advancing Translational Sciences, and R21EB030209 from the National Institute of Biomedical Imaging and Bioengineering of the National Institutes of Health, National Institutes of Health, USA. The content is solely the responsibility of the authors and does not necessarily represent the official views of the National Institutes of Health. This research has been partially funded through the generous support of Herbert and Florence Irving/the Irving Trust.